\documentclass[conference]{IEEEtran}
\usepackage[scr=rsfs]{mathalpha}

\usepackage{graphicx, amsfonts,amsmath , boldline, epstopdf,amssymb }
\usepackage{tabu, multirow,multicol ,booktabs ,setspace ,algcompatible ,mathrsfs,dsfont, array, boldline, makecell, booktabs }
\usepackage{soul}
\usepackage{amssymb}
\usepackage[linesnumbered,lined,boxed,commentsnumbered,ruled,longend]{algorithm2e}
\usepackage{adjustbox, stackengine, color, colortbl, epsfig,cite, cases, enumitem, mathtools, tikz, verbatim}
\newcolumntype{?}{!{\vrule width 1.2pt}} 
\usepackage[caption = false]{subfig}
\usepackage[utf8]{inputenc}
\makeatletter

\begin{document}

\title{\vspace{-.1cm} Economic Viability of the Energy-Water-Hydrogen Nexus for Power System Decarbonization \\
\thanks{Identify applicable funding agency here. If none, delete this.}
}

\author{\IEEEauthorblockN{ Mostafa Goodarzi}
\IEEEauthorblockA{\textit{Electrical and Computer Engineering} \\
\textit{University of Central Florida}\\
Orlando, USA \\
mostafa.goodarzi@ucf.edu}
\and
\IEEEauthorblockN{ Qifeng Li}
\IEEEauthorblockA{\textit{Electrical and Computer Engineering} \\
\textit{University of Central Florida}\\
Orlando, USA \\
Qifeng.Li@ucf.edu}}

\maketitle

\begin{abstract}
This paper aims to evaluate the economic viability of energy-water-hydrogen (EWH) nexus as a new solution for reducing carbon emissions from power systems. The urgency around climate change emphasizes the pressing need to mitigate carbon emissions, especially from the electricity sector, which accounts for a significant portion of total emissions in the US. In response, incorporating more renewable energy sources (RESs) and green hydrogen—created through water electrolysis and RES—stands out as a crucial strategy to combat climate challenges. We delve into various aspects of the EWH nexus, including carbon emissions from different power plants, capturing these emissions, and potential options for their reuse or storage. This paper involves modeling different sections of the EWH nexus and conducting an economic analysis across scenarios in power plants to determine optimal water supply methods, suitable chemical products for carbon reuse, and an appropriate carbon emission penalty to encourage emission reduction through the EWH nexus. The results indicate that reusing captured carbon emissions emerges as the most beneficial option across all power plant types. This finding underscores the potential of carbon reuse as a pivotal strategy within the EWH nexus framework for addressing carbon emissions.
\end{abstract}


\allowdisplaybreaks
\section{Introduction}   \label{sec: Intro}
The heightened urgency surrounding climate change has emphasized the critical need to address carbon emissions, particularly from the power sector, which accounts for a quarter of the total emissions in the US. Integrating a higher proportion of renewable energy sources (RESs) has emerged as a potent strategy aligned with the nation's commitment to global climate agreements such as the Paris Agreement. However, the unpredictable nature of RESs presents challenges in scheduling the power system. Additionally, RES integration into the power grid faces limitations due to grid hosting capacity \cite{samet2020deep}. Consequently, the complete replacement of fossil fuel power plants with RESs is currently unfeasible.

An alternative approach involves employing a carbon capture and storage system (CCSS) in fossil fuel power plants to reduce carbon emissions. Trapping and separating $CO_2$ from other gases, moving the captured $CO_2$ to a storage area, and storing it away from the atmosphere are all parts of the CCSS process \cite{elhenawy2020metal}. Although storage of captured carbon is common, its safety and potential environmental impact remain uncertain. This concern can be addressed by reusing captured carbon by mixing it with hydrogen to create chemical products. This paper evaluates the options for storing or reusing captured carbon.

The power-to-X (P2X), which uses RES to produce hydrogen and make chemical products \cite{burre2020power}, has recently attracted significant attention as a method of decarbonization \cite{varela2021modeling}. The P2X system relies heavily on hydrogen, which can be derived from fossil fuels or from RESs known as green hydrogen. Green hydrogen, produced from RES through the electrolysis process, can reduce greenhouse gas emissions by 94\% compared to hydrogen derived from natural gas processes\cite{ghandehariun2016life}. In addition to electricity, water is a crucial component in hydrogen production through electrolysis. If all current hydrogen production is assumed to be from water electrolysis, it will result in an estimated consumption of around 70 million metric tons of water \cite{yue2021hydrogen}. The significance of water in electrolysis-based hydrogen production highlights the need for exploring the integration of water systems with green hydrogen production, an area that lacks a comprehensive study. This paper proposes a novel concept of the Energy-Water-Hydrogen (EWH) nexus as a promising solution to reduce power system carbon emissions.

While the P2X system has undergone detailed study, the broader concept of the EWH nexus remains relatively unexplored. In our research, we delve into an economic analysis of the EWH nexus from the perspective of P2X to address various open questions. These include determining the most suitable water supply methods, exploring diverse scenarios concerning the storing and reutilization of captured carbon, assessing different chemical products, and establishing a reasonable pricing value for carbon emission penalties. To fill this gap, we propose a mathematical model for the EWH nexus and conduct an economic analysis, aiming to provide insights and answers to these critical questions.

The rest of this paper is organized as follows. Section \ref{sec: EWHN} introduces the novel EWH nexus concept. In Section \ref{sec: EWHNmodel}, the detailed model of EWH nexus aimed to reduce carbon emissions from conventional power plants is presented. Section \ref{sec: Results} covers the economic analysis and results. Finally, conclusions and future work are drawn in Section \ref{sec: conclusions}.

\section{Energy-Water-Hydrogen Nexus}  \label{sec: EWHN}

\subsection{Illustrating the EWH Nexus Configuration}
The EWH nexus involves utilizing clean, cost-effective, and uncontrollable RES for water electrolysis, generating controllable energy in the form of hydrogen. The use of hydrogen spans many sectors, such as transportation and industry. It can also be fed back into the power system through fuel cells to have a controllable power source. Power, water, and hydrogen systems are the three core sections of the EWH nexus, which cooperate together. In the water section, power is essential for various operations like water treatment, desalination, and pumping, while water plays a critical role in the power system, particularly for cooling purposes. Moreover, the hydrogen production segment requires both power and water to facilitate the electrolysis process for hydrogen generation.

\subsection{EWH Nexus for Reducing Carbon Emission}
\begin{figure}[!t]
  \centering
\includegraphics[width=0.489\textwidth]{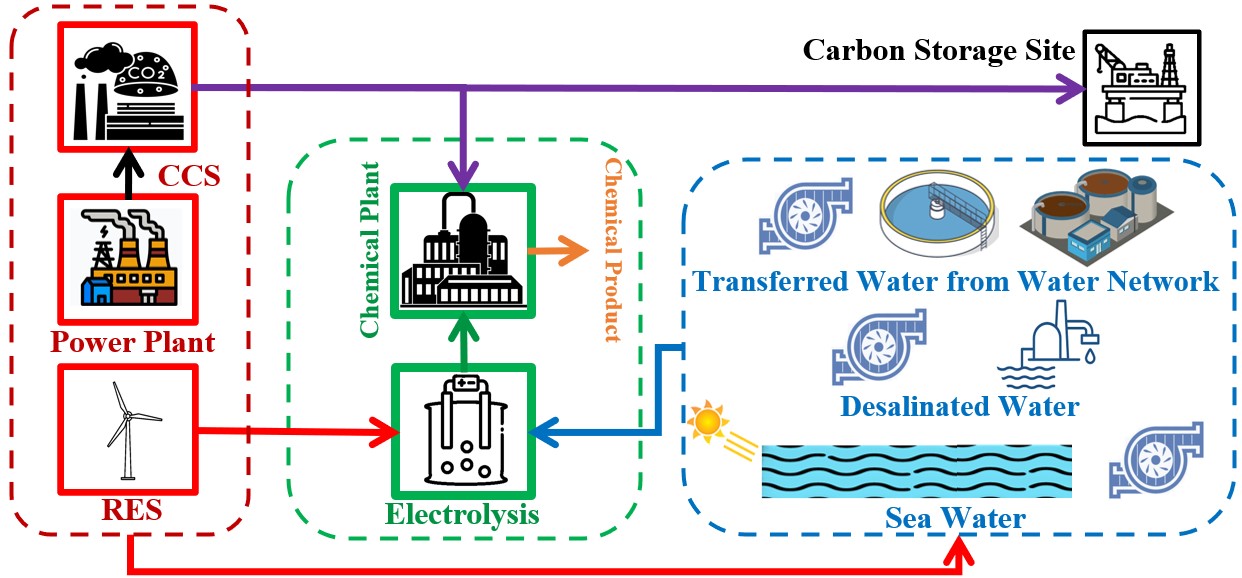}
 \centering
 \vspace{-.3cm}
\caption{\footnotesize A typical EWH nexus for decarbonization of fossil fuel power plant: \textcolor{red}{\textbf{$\rightarrow$}} power, \textcolor{blue}{\textbf{$\rightarrow$}} water, \textcolor{black}{\textbf{$\rightarrow$}} flu gas, \textcolor{purple}{\textbf{$\rightarrow$}} carbon, \textcolor{green}{\textbf{$\rightarrow$}} hydrogen, \textcolor{orange}{\textbf{$\rightarrow$}} chemical product, \textcolor{red}{\textbf{-\,-\,-}} energy section, \textcolor{blue}{\textbf{-\,-\,-}} water section, and \textcolor{green}{\textbf{-\,-\,-}} hydrogen section.}
  \label{Fig: EWHNCarbon}  
        \vspace{-.5cm}
\end{figure}

We propose to use an EWH nexus to reuse the captured carbon obtained from the carbon capture system (CCS). Fig. \ref{Fig: EWHNCarbon} illustrates the EWH nexus to reduce the carbon emissions of conventional power plants.
The targeted conventional power plant for decarbonization is equipped with CCS to capture carbon emissions. These emissions can be stored or combined with hydrogen from the EWH nexus to produce a chemical product. In the power section of the EWH nexus, wind energy is utilized for water electrolysis. Within the water section, three types of water resources are considered: transferring water from the network and using tap water, establishing an additional desalination plant for purifying seawater, and directly employing seawater for hydrogen production using solar energy. The hydrogen section involves water electrolysis and a chemical plant to combine the hydrogen with carbon dioxide from CCS and create a chemical product. A mathematical model of this system is further explored in the following section, aiming to assess the economics of an EWH nexus coupled with a CCS-equipped conventional power plant to reduce carbon emissions.



\section{Mathematical Modeling}  \label{sec: EWHNmodel}
\subsection{Power Section}
There are many conventional power plants that emit carbon dioxide into the atmosphere, with coal, gas, and biomass power plants ranking as the top three emitters, releasing 820, 490, and 230 $g/kWh$ of carbon, respectively  \cite{eiaFAQ}. These plants can reduce carbon emissions by implementing CCS, available in three types: post-combustion, pre-combustion, and oxyfuel combustion \cite{thiruvenkatachari2009post}. Our model proposes the post-combustion type due to its adaptability to new and existing power plants. The captured $CO_2$ can be transported to a storage facility through pipelines, ships, or trucks. While shipping proves cost-effective for long-distance $CO_2$ transport, pipelines typically offer the most cost-effective solution in many regions \cite{smith2021cost}. Therefore, pipelines are our preferred method of transportation. Although storing captured carbon is common today, its process is not thoroughly tested, raising concerns about marine life safety. Hence, we suggest the adoption of reusing as an alternative method to manage captured carbon within an EWH nexus. The following formulations present a mathematical model detailing the operational and capital cost of a CCSS.
\begin{subequations} \label{eq: CCSS}
\begin{align}
&c^\textrm{ccss} = ((1-\beta) c^\textrm{cts}  + c^\textrm{ccs}) \overline{C},\label{eq: CCSS1}\\
&\gamma^\textrm{ccss} = \sum_{t} ((1-\beta) c_t  r^\textrm{cts} + c_t r^\textrm{ccs}), \label{eq: CCSS2}
\end{align}
\end{subequations} 
where (\ref{eq: CCSS1}) represents the cost to build a CCS and construct the pipeline, and (\ref{eq: CCSS2}) shows the operational cost of CCSS. $c^\textrm{ccss}$, $\gamma^\textrm{ccss}$, $c^\textrm{cts}$, $r^\textrm{cts}$, $c^\textrm{ccs}$, and $r^\textrm{ccs}$ are the capital and operational cost of CCSS, and the unit capital and operational cost transferring system, and CCS, respectively. $c_t$ shows the carbon that is captured by the CCS at time $t$ and $\overline{C}$ represents carbon emissions produced by the power plant when operating at maximum capacity. We propose reusing a portion of captured carbon as shown in (\ref{eq: CCSS}) with $\beta$. 

Meeting the energy demands for water electrolysis is a key challenge in the EWH nexus. Hydrogen production via electrolysis requires approximately 50-55 $kWh/kg$ of energy. Wind energy is suggested in our model to meet this energy. Eq (\ref{eq: P_EWHN}) shows the capital cost of the power section.
\begin{align}  \label{eq: P_EWHN}
c^\textrm{p} = \frac{c^\textrm{wind}  \xi^\textrm{p}\overline{H}}{0.423},
\end{align}
where 42.3\% shows the annual net capacity factor for a wind farm\cite{stehly20212020}. $c^\textrm{p}$ and $c^\textrm{wind}$ represent the capital cost of the power section and the unit capital cost for constructing a wind farm, respectively. $\xi^\textrm{p}$ shows the unit required energy for the hydrogen system and $\overline{H}$ represents hydrogen produced by the hydrogen section when operating at maximum capacity.

\subsection{Water Section}
Water serves as an environmentally friendly source of hydrogen without carbon emissions. Various water supply options for hydrogen production are being considered in the proposed EWH nexus. These options include the transfer of tap water from existing water networks, the construction of a desalination facility near the power plant, and the utilization of solar energy alongside seawater to generate water suitable for electrolysis.
Delivery of tap water to the EWH nexus can vary based on the geographic location of the power plant. Alternatively, it can be produced locally via desalination plants near the power plant, or it can be transferred from existing water networks. Since a large number of power plants are located near coastal regions, seawater desalination is a feasible option for EWH nexus. We employ piece-wise linearization to estimate the power demand of the reverse osmosis desalination as follows \cite{mohammadi2019coordinated,goodarzi2022evaluate}:
\begin{align}
p^\mathrm{des}_t = e^\mathrm{des}_k  f_t, & & & 0.25(k-1) \overline{W} \leq f_t \leq 0.25k \overline{W},
\label{eq:DesPower2}
\end{align}
where $k \in \{1,2,3,4\}$. $p^\mathrm{des}_t$, $e^\mathrm{des}_k$, $f_t$, and $\overline{W}$ show total desalination power, desalination power constant, water flow at time $t$, and maximum water production capacity, respectively. 
To transport water from the water network to the power plant at a suitable head pressure, the water pump needs to elevate the pressure to overcome the losses experienced along the pipe. We assume that the head pressure within the water distribution system is equivalent to the head pressure at the power plant, making the head gain equal to the head loss along the pipe. The power of the pump is modeled as follows \cite{goodarzi2021fast}:
 \begin{subequations}\label{eq:headloss}
 \begin{align}
&y^\mathrm{G}_{t}= r^\mathrm{w}f_{t}^2, \label{eq:headloss1}\\
&\eta p^{\textrm {pump}}_{t} = 2.725 y^\textrm{G}_t  f_t, \label{eq:pump2}
\end{align}
\end{subequations}
where $y^\mathrm{G}_{t}$ is water head at time $t$. $r^\mathrm{w}$ and $p^{\textrm {pump}}$ display the head loss coefficient and power of the pump.
The seawater scenario includes solar energy and seawater to produce hydrogen. Solar energy is commonly employed through two distinct processes for hydrogen production: solar-powered water electrolysis and direct solar water splitting. Solar-powered water electrolysis entails relatively minimal energy, capital, and operational costs \cite{sedayevatan2023uncertainty} compared to water splitting. Consequently, direct seawater splitting lacks substantial advantages and presents considerable drawbacks \cite{hausmann2021direct}. The conversion of seawater directly into hydrogen is not currently a viable option, but that may change in the future. Therefore, we only consider the solar-powered water electrolysis method. The general formulation for the water system is as follows: 
\begin{subequations}
	\begin{align}
&c^\textrm{w} = \overline{W} (\alpha_1 c^\textrm{des} + \alpha_2 c^\textrm{tw} d + \alpha_3 c^\textrm{sw}), \label{eq:water3_2}\\
&\gamma^\textrm{w} = (1-\alpha_3) \sum_{t} (\alpha_1 r_t^\textrm{des}  p_t^\textrm{Des} + \alpha_2 r_t^\textrm{pump} p_t^\textrm{pump} ),\label{eq:water4_2}\\
& \alpha_1 + \alpha_2 + \alpha_3 = 1, \label{eq:water4_3}
	\end{align}
\end{subequations}
Where (\ref{eq:water3_2}) and (\ref{eq:water4_2}) represent the capital and operational cost of the water system in the proposed model. $\alpha$ is a binary variable that relates to the choice between employing desalination, transferring water, or adopting a scenario involving solar-powered water electrolysis. $c^\textrm{des}$, $c^\textrm{tw}$, and  $c^\textrm{sw}$ represent the unit capital cost of desalination, transferring water, and solar-powered water electrolysis, respectively. $d$, $r_t^\textrm{des}$, and $r_t^\textrm{pump}$ display the distance between the water network and electrolysis, and the operational cost of the water desalination and water pump, respectively. Equation (\ref{eq:water4_3}) guarantees the use of only one method to supply water for the EWH nexus.

\subsection{Hydrogen Section}
The hydrogen section encompasses the process of water electrolysis for hydrogen production and a chemical facility for generating a chemical product, such as methane. The following mathematical formulations describe the capital cost of electrolysis which is the only equipment of the hydrogen section:
\begin{align}
c^\textrm{h} = \xi^\textrm{h} \beta \overline{C} c^\textrm{we}. \label{eq_H2_2}
\end{align}
We consider a group of electrolysis units with a capital cost of $c^\textrm{we}$. The parameter $\xi^\textrm{h}$ represents the amount of hydrogen needed to reuse 1 $kg$ of captured carbon.

The combination of green hydrogen with captured carbon allows for the creation of diverse chemical outputs through the synthesis of hydrogen and captured carbon dioxide, offering a range of potential products, such as methane ($CH_4$), ethanol ($C_2H_6O$), and methanol ($CH_3OH$), for commercial use. 
Hydrogen generated through water electrolysis can be combined with captured carbon to produce methane as a primary component of natural gas. Based on the chemical equation for this process ($4H_2 + CO_2 \rightarrow CH_4 + 2H_2O$), we need $182$ $g$ of hydrogen to recycle $1$ $kg$ of $CO_2$, which can be produced from $1.64$ $liters$ of water.
Methanol stands out as a highly promising product within the P2X concept due to its extensive applications in both the chemical and energy industries, resulting in the creation of high-value products \cite{kourkoumpas2016implementation}. As per the chemical equation describing this process ($CO_2 + 3H_2 \rightarrow CH_3OH + H_2O$), it's apparent that to recycle $1$ $kg$  of $CO_2$, $137.4$ $g$ of hydrogen is required. 
Ethanol is a renewable fuel made from various plant materials collectively known as biomass. Based on the chemical equation for this process ($2CO_2 + 6H_2 \rightarrow CH_3OCH_3 + 3H_2O$), we need $182$ $g$ of hydrogen to recycle $1$ $kg$ of $CO_2$, which can be produced from $1.64$ $liters$ of water. The proposed model incorporates revenue generated from selling the chemical product.
\begin{align}  \label{eq:gas}
&\gamma^\chi = - \sum_{t} \rho^\chi \xi^\chi \beta c_t,
\end{align}
where $\rho^\chi$ represents the unit revenue of the chemical product, while $\xi^\chi$ signifies the quantity of the chemical product obtained by reusing 1 $kg$ of $CO_2$. We assume the existence of a chemical factory dedicated to chemical products. Hence, we do not consider the capital cost of the chemical section.

\subsection{Total Cost of the EWH nexus}
The total cost of the proposed EWH nexus system comprises both operational and capital costs, outlined as follows:
\begin{align}  \label{eq:totcost}
&C^\textrm{tot} = \frac{c^\textrm{ccss} + c^\textrm{p} + c^\textrm{w}}{365N} (1+\lambda)^{N-1} + \gamma^\textrm{ccss} +\gamma^\textrm{w} + \gamma^\chi,
\end{align}
where $\lambda$ shows the interest rate for the capital costs, while $N$ represents the project's duration over which we assess the return on investment.

\section{economic analysis and results}  \label{sec: Results}
This section conducts a comprehensive economic analysis on the proposed EWH nexus under different scenarios of water resources and carbon reuse/storage. A reasonable carbon emission penalty rate to incentivize conventional power plants to minimize their carbon emissions will also be explored. 

\subsection{Parameter Setting}
The model considers a consistent electricity price of $\$0.25/kWh$ and a natural gas price of $\$1400/ton$. Notably, the prices for chemical products are specified as $\$616/ton$ for methanol, and $\$493/ton$ for ethanol. $r^\textrm{Cap}_\textrm{CTS}$ varies from 1 to 6 $\$/ton$ depending on the daily carbon mass flow rate, $d = 250 km$, $r^\textrm{cts} = 15\$/ton$, $r^\textrm{ccs} = 45\$/ton$, $c^\textrm{wind} = 1030$, $c^\textrm{des} = 0.2 M\$/m^3$, $c^\textrm{tw} = 160 \$/m$, and $\eta = 90\%$. \cite{Methanolprice,SEC,smith2021cost,herzog2009carbon,stehly20212020}. This paper analyzes three power plants: coal, natural gas, and biomass, all with 500 MW capacity. Table \ref{tab: Cemission2} displays the carbon emissions, hydrogen requirements for recycling all of these emissions, water needs for hydrogen production, and chemical production across different power plants and chemical products.

\begin{table}[!b]
\vspace{-.5cm}
\centering
\footnotesize
\captionsetup{labelsep=space,font={footnotesize,sc}}
\caption{\textrm{Specification of EWH nexus Sections for Diverse Power Plants}}
\vspace{-.2cm}
\footnotesize
\begin{tabular}{ccccc}
\hline \hline
Power Plant &  Carbon  & Required  &Required  &Chemical  \\ 
 &   Emission &  Hydrogen & Water & Production \\
 &   $(ton/h)$ &  $(ton/h)$ & $(m^/h)$ & $(ton/h)$ \\
\midrule
\multicolumn{5}{c}{\footnotesize\textbf{ Chemical Product: Methane}}\\
Biomass &  115 &  21 &  188 & 42 \\ 
Natural Gas &  245 &  45 &  401 & 89  \\ 
Coal &  410 &  75 &  671 & 149  \\ 
\midrule
\multicolumn{5}{c}{\footnotesize\textbf{ Chemical Product: Methanol}}\\
Biomass &  115 &  16 &  142 & 84 \\ 
Natural Gas &  245 &  34 &  303 & 178  \\ 
Coal &  410 &  56 &  501 & 298  \\ 
\midrule
\multicolumn{5}{c}{\footnotesize\textbf{ Chemical Product: Ethanol}}\\
Biomass &  115 &  16 &  142 & 60 \\ 
Natural Gas &  245 &  34 &  303 & 128  \\ 
Coal &  410 &  56 &  501 & 214  \\
\midrule
\midrule
\end{tabular}
\label{tab: Cemission2}
\end{table}

\subsection{The best type of water resources for EWH nexus}
\begin{figure}[!t]
  \centering
{\includegraphics[width=.4\textwidth]{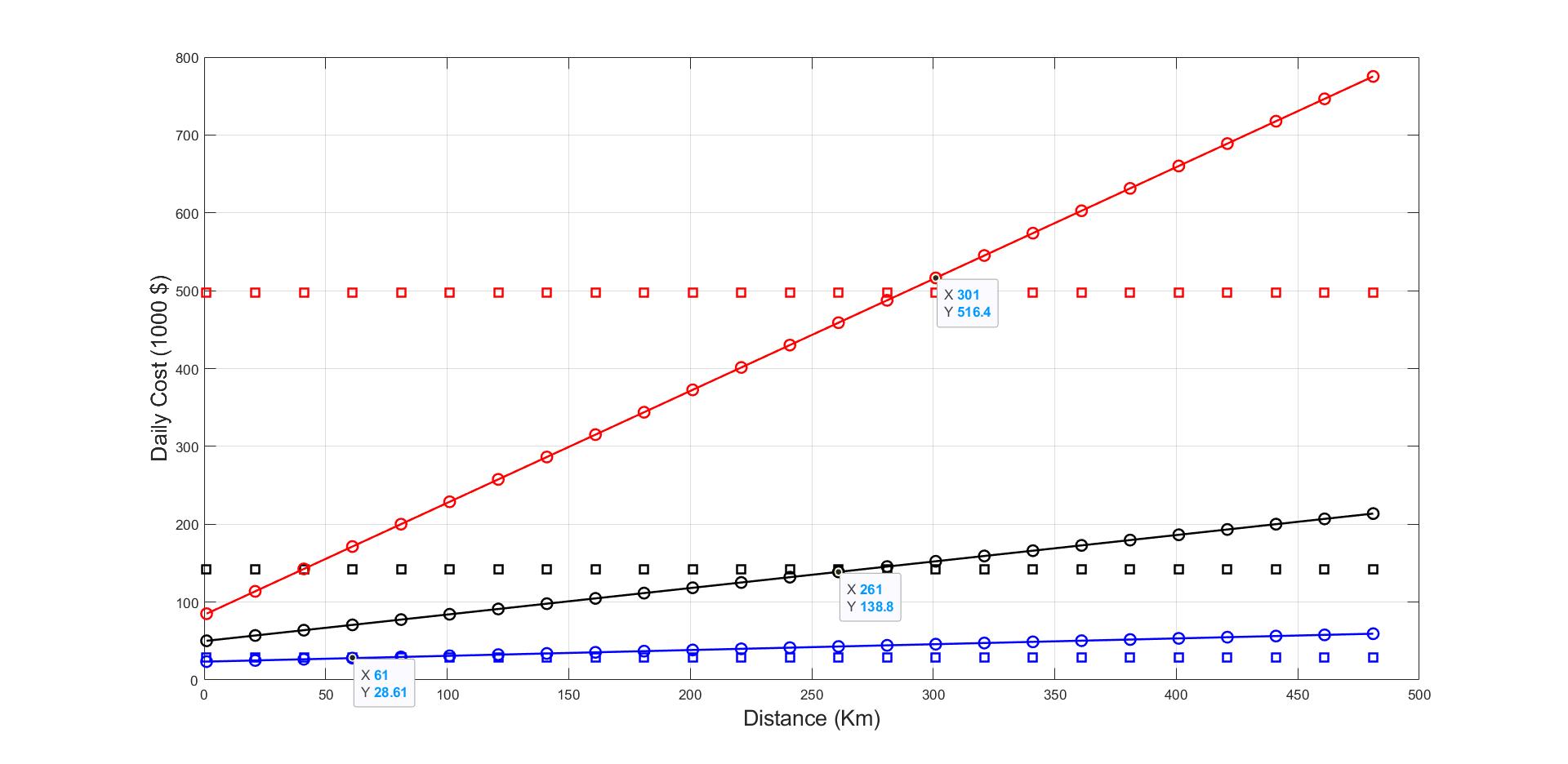}}
 \centering
 \vspace{-.3cm}
    \caption{Comparison of the costs between two distinct water systems for various power plants $\square$: Construction of desalination, $\circ$: Construction of pipe. \textcolor{blue}{-o-} Biomass, \textcolor{black}{\textbf{-o-}}: Natural Gas, \textcolor{red}{\textbf{-o-}}: Coal.}
  \label{pic:Des_pipe}
 \vspace{-.5cm}
\end{figure}

\begin{figure}[!b]
\vspace{-.6cm}
  \centering
  \subfloat[]{\includegraphics[width=0.16\textwidth]{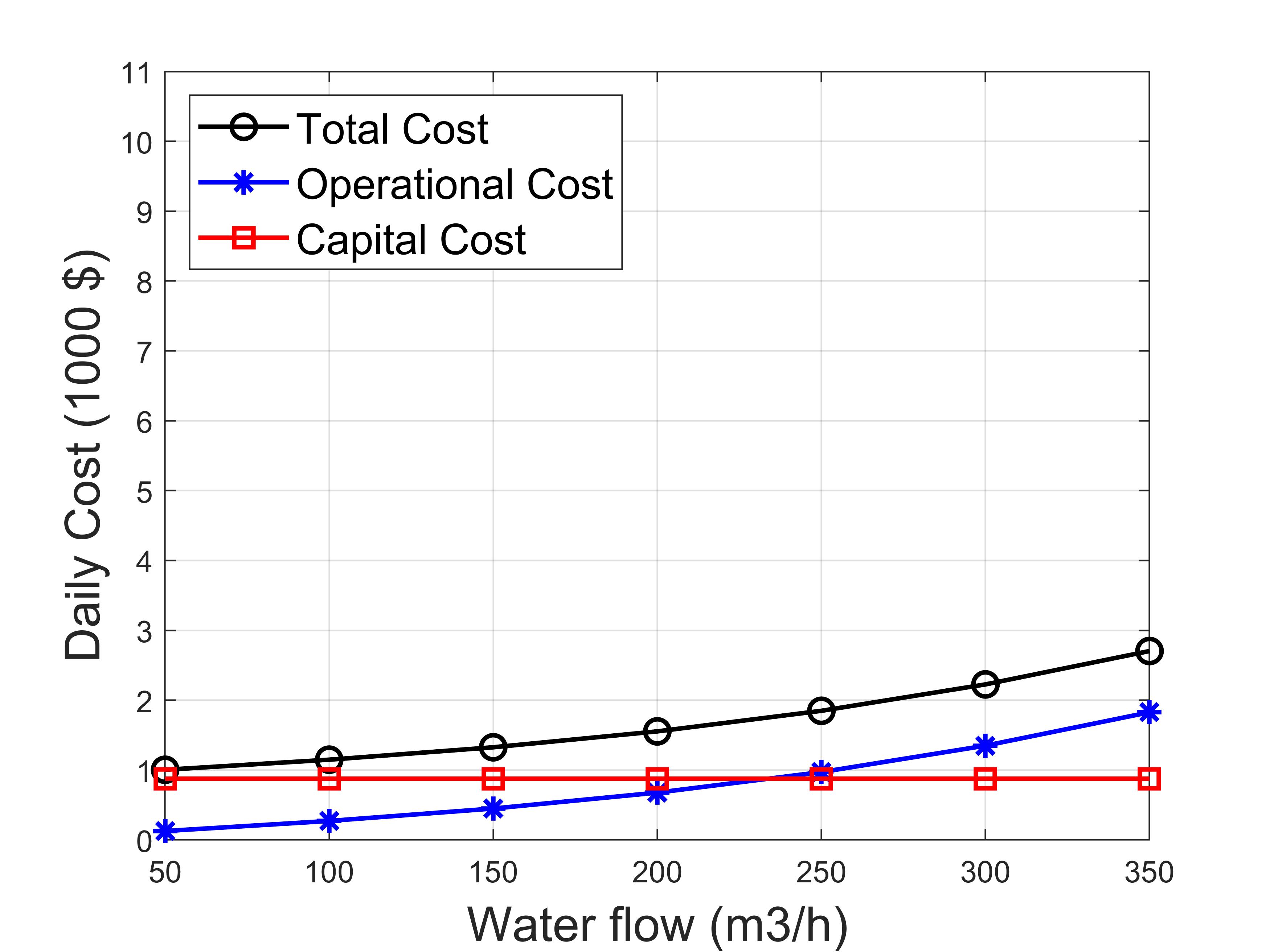}\label{Pic:dis60km}}
      \subfloat[]{\includegraphics[width=0.16\textwidth]{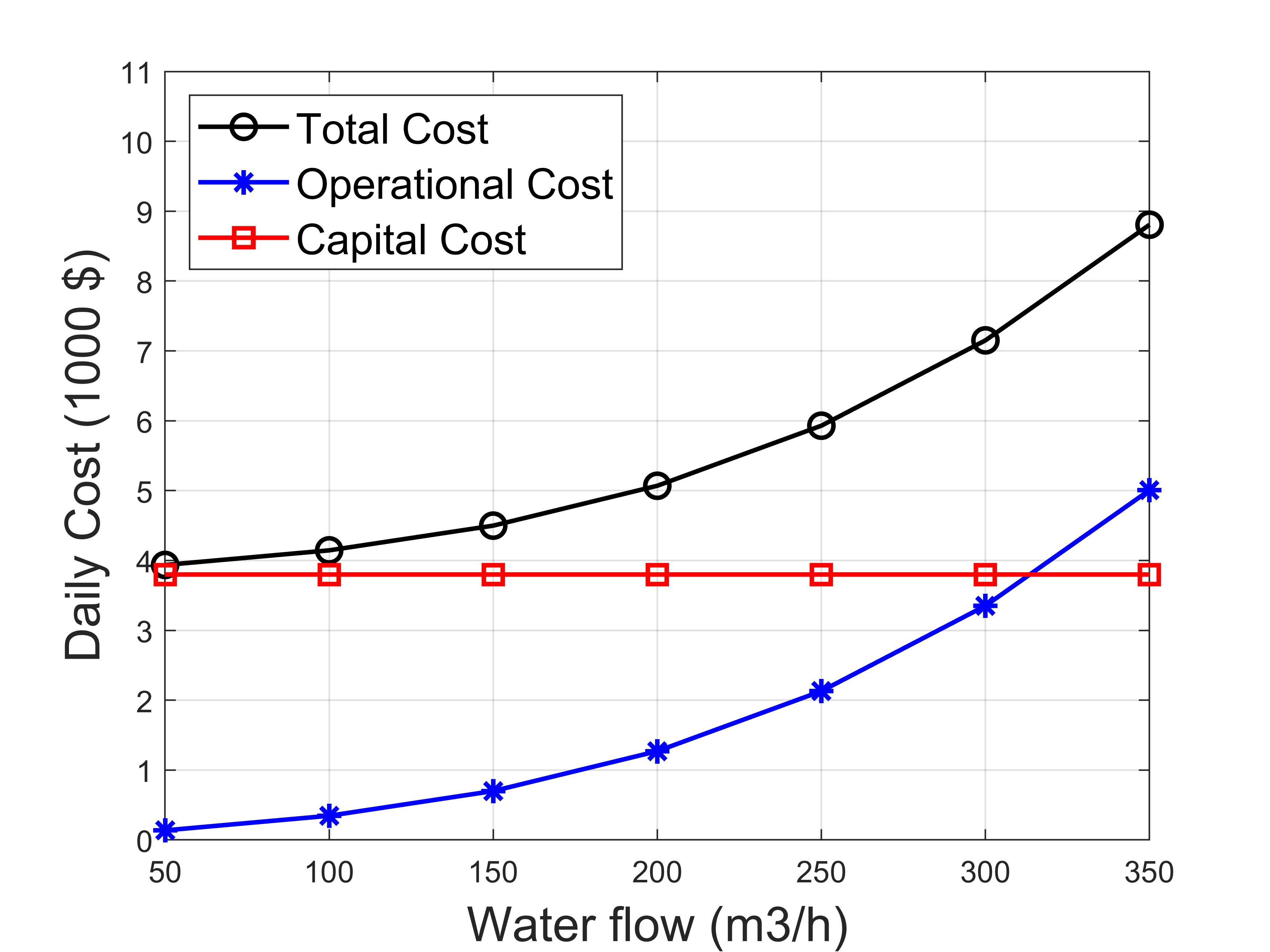}\label{Pic:dis260km}}
    \subfloat[]{\includegraphics[width=0.16\textwidth]{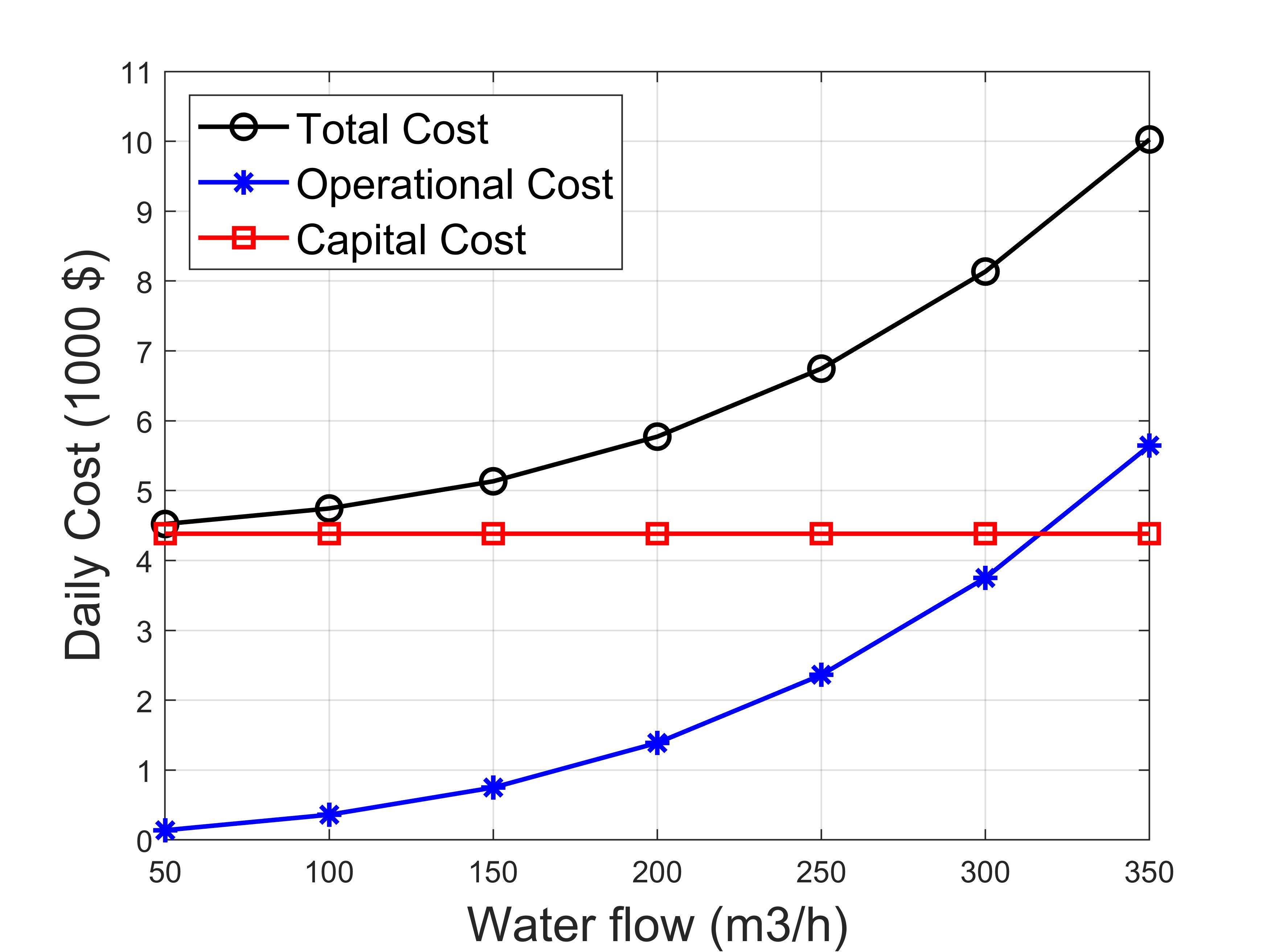}\label{Pic:dis300km}}
    \caption{Daily cost for transferring different flows of water over different distances: a) 60 $km$  b) 260 $km$ c) 300 $km$}
    \label{Pic:dis}
\end{figure}

This section evaluates two water supply options for the proposed EWH nexus. One option involves establishing a desalination plant close to the power station, while the second one involves laying pipes to transport water from the existing water network to the power plant. The first option necessitates factoring in the construction costs of the desalination plant and ongoing operational expenses related to purchasing power to satisfy the desalination demand. Conversely, the cost of the second system will fluctuate depending on the distance between the power plant and the water network. The comparison in Fig. \ref{pic:Des_pipe} illustrates the cost analysis of these two different water supply options across various power plants, considering the distance between the power plant and the water system. This figure demonstrates that when the distance between the water network and the power plant exceeds 61 km for biomass, 261 km for gas, and 301 km for coal power plants, it becomes more advantageous to construct a new water desalination system rather than transferring water from the existing water network.
Fig. \ref{Pic:dis} displays the capital cost, operational cost, and total cost associated with water transfer from the water network to the power plant for these three distances. Considering potential alterations in the capacity of the power plants, the required water quantity for the electrolysis process might fluctuate. 

\subsection{Reusing $CO_2$ versus Transferring to Storage Locations}
In this section, we compare the economic viability between 'reusing' captured carbon and 'storing' it. The paper assumes that power plants are located near coastal areas due to their proximity to the sea. Consequently, the desalination plant facilitates the necessary water supply for water electrolysis. In the context of carbon transfer, the pipeline option is chosen due to its recognized cost efficiency.

The daily costs of different power plants under three scenarios—storing all carbon, storing 50\% of captured carbon, and reusing all captured carbon—have been studied. The most cost-effective scenario involves reusing all captured carbon by converting it into methane. Furthermore, we calculate the necessary increase in the selling price of electricity generated by a power plant to offset the costs incurred by implementing a CCS. For instance, in the biomass power plant, to cover all the expenses stemming from integrating the CCSS, power plant owners should raise the electricity selling price by $\$0.41/kWh$.
We also calculate the carbon penalty. This information helps power plant owners to determine the economic threshold for emitting carbon into the atmosphere and paying the penalty. Table \ref{tab: Costprice} summarizes comprehensive data for three types of power plants under different scenarios. For example, a methane-based EWH nexus could generate revenue when reusing all carbon emissions from a power plant.

\begin{table}[!t]
\centering
\caption{\textrm{{Cost and Price Impact: $CO_2$ Reusing via EWH nexus}}}
\vspace{-.2cm}
\footnotesize
\begin{tabular}{ccccccc}
\hline \hline
\multirow{2}{*}{Power Plant} &  \multicolumn{2}{c}{Daily cost}  & \multicolumn{2}{c}{Increased Price}  &\multicolumn{2}{c}{Carbon Penalty}   \\ 
 &  \multicolumn{2}{c}{($M\$/day$)}  & \multicolumn{2}{c}{($\$/kWh$)}  &\multicolumn{2}{c}{($\$/tone$)}   \\ 

\midrule
\multicolumn{7}{c}{\footnotesize\textbf{ All Captured Carbon Storing}} \vspace{.1cm}\\
Biomass &   \multicolumn{2}{c}{0.207}  & \multicolumn{2}{c}{0.41}  &\multicolumn{2}{c}{75.09} \\ 
Natural Gas &   \multicolumn{2}{c}{0.395}  & \multicolumn{2}{c}{0.79}  &\multicolumn{2}{c}{67.19}  \\ 
Coal &  \multicolumn{2}{c}{0.633}  & \multicolumn{2}{c}{1.27}  &\multicolumn{2}{c}{64.38}  \\ 
\midrule
\multicolumn{7}{c}{\footnotesize\textbf{ Reusing Captured Carbon for Chemical Product: Methane}}\vspace{.1cm}\\
& half  & all  & half   & all  & half   & all    \\ 
Biomass & 0.0980 & -0.0097 &  0.2 & -0.02 & 35.52 & -3.52 \\ 
Natural Gas &  0.1640 & -0.0626 &  0.33 & -0.12 &27.78 & -10.65  \\ 
Coal &  0.2507 & -0.1225 &  0.50 & -0.24 & 25.47 &-12.45  \\ 
\midrule
\multicolumn{7}{c}{\footnotesize\textbf{Reusing Captured Carbon for Chemical Product: Methanol}}\vspace{.1cm}\\
& half  & all  & half   & all  & half   & all    \\ 
Biomass &  0.1405 & 0.0737 &  0.28 & 0.15 & 50.92 & 26.71 \\ 
Natural Gas & 0.2542 & 0.1165 &  0.51 & 0.23 & 43.22 & 19.81  \\ 
Coal &  0.3977 & 0.1748 &  0.79 & 0.35 & 40.41 &17.76  \\ 
\midrule
\multicolumn{7}{c}{\footnotesize\textbf{Reusing Captured Carbon for Chemical Product: Ethanol}}\vspace{.1cm}\\
& half  & all  & half   & all  & half   & all    \\ 
Biomass &  0.2129 &0.2185 &  0.43 &0.44 & 77.15 &79.17 \\ 
Natural Gas &  0.4084 & 0.4250 &  0.82 &0.85 &69.46 &72.27  \\ 
Coal &  0.6558 & 0.6910 &  1.31 & 1.38 & 66.64 &70.22  \\
\midrule
\midrule
\end{tabular}
\label{tab: Costprice}
\vspace{-.5cm}
\end{table}

\subsection{Carbon Emission Penalty Policy}
This section establishes a feasible carbon emissions penalty, providing power plant owners with the choice to pay the penalty or seek options to reuse or transport the captured carbon. Decision-makers need to determine a rational penalty rate to motivate power plant owners to adopt the EWH nexus and diminish carbon emissions. 
Based on Table \ref{tab: Costprice}, it becomes apparent that a penalty exceeding $\$75.09/tone$ serves as a compelling incentive for power plant owners to invest in CCSS and store all carbon emissions. The scenario changes when considering the adoption of an EWH nexus, where the reuse of all captured carbon to reduce ethanol requires a carbon penalty price exceeding $\$79.19/tone$. This threshold will be $\$26.71/tone$ for an EWH nexus with methanol production. The proposed EWH nexus will prove economically viable if it includes a methane chemical plant, which will cover all capital costs and generate additional revenue. As a result, any carbon emission penalty price becomes a motivating factor for owners of conventional power plants to enhance their facilities through the incorporation of an EWH nexus.

\section{Conclusions and Future Work}\label{sec: conclusions}

This paper presents the EWH nexus as a novel and innovative engineering approach aimed at amalgamating conventional power plants, CCSS, RES, water systems, hydrogen systems, and chemical systems to effectively reduce carbon emissions in the power sector. The discussion covers various critical aspects, including carbon emissions from diverse power plants. It also examines the process of capturing and storing these emissions and explores potential methods for their reuse. Furthermore, an extensive economic analysis is conducted across various segments of the EWH nexus. The discussion focused on different water supply methods adapted to specific locations, proposing the most viable options. Among the three studied chemical products—methane, methanol, and ethanol—as potential candidates for reusing captured carbon, methane emerged as the most promising option. It is currently common for power plants with CCS to transfer the captured carbon to storage locations, but according to our findings, using this carbon in conjunction with green hydrogen reduces costs and, in some cases, generates revenue as well. Therefore, implementing EWH nexus in conventional power plants will reduce carbon emissions and costs.

Future works involve extending the proposed model to the microgrid at the distribution level to obtain optimal operation of the distribution-level EWH nexus to enhance RES and reduce carbon emissions. Additionally, the influence of control strategies on maintenance costs and equipment lifespan within the EWH nexus can be explored.

\bibliographystyle{IEEEtran}	
\bibliography{Main}

\begin{thebibliography}{10}
\providecommand{\url}[1]{#1}
\csname url@samestyle\endcsname
\providecommand{\newblock}{\relax}
\providecommand{\bibinfo}[2]{#2}
\providecommand{\BIBentrySTDinterwordspacing}{\spaceskip=0pt\relax}
\providecommand{\BIBentryALTinterwordstretchfactor}{4}
\providecommand{\BIBentryALTinterwordspacing}{\spaceskip=\fontdimen2\font plus
\BIBentryALTinterwordstretchfactor\fontdimen3\font minus \fontdimen4\font\relax}
\providecommand{\BIBforeignlanguage}[2]{{%
\expandafter\ifx\csname l@#1\endcsname\relax
\typeout{** WARNING: IEEEtran.bst: No hyphenation pattern has been}%
\typeout{** loaded for the language `#1'. Using the pattern for}%
\typeout{** the default language instead.}%
\else
\language=\csname l@#1\endcsname
\fi
#2}}
\providecommand{\BIBdecl}{\relax}
\BIBdecl

\bibitem{samet2020deep}
H.~Samet, S.~Ketabipour, M.~Afrasiabi, S.~Afrasiabi, and M.~Mohammadi, ``Deep learning forecaster-based controller for svc: Wind farm flicker mitigation,'' \emph{IEEE Transactions on Industrial Informatics}, vol.~18, no.~10, pp. 7030--7037, 2020.

\bibitem{elhenawy2020metal}
S.~Elhenawy, M.~Khraisheh, F.~AlMomani, and G.~Walker, ``Metal-organic frameworks as a platform for co2 capture and chemical processes: Adsorption, membrane separation, catalytic-conversion, and electrochemical reduction of co2,'' \emph{Catalysts}, vol.~10, no.~11, p. 1293, 2020.

\bibitem{burre2020power}
J.~Burre, D.~Bongartz, L.~Br{\'e}e, K.~Roh, and A.~Mitsos, ``Power-to-x: between electricity storage, e-production, and demand side management,'' \emph{Chemie Ingenieur Technik}, vol.~92, no. 1-2, pp. 74--84, 2020.

\bibitem{varela2021modeling}
C.~Varela, M.~Mostafa, and E.~Zondervan, ``Modeling alkaline water electrolysis for power-to-x applications: A scheduling approach,'' \emph{International journal of hydrogen energy}, vol.~46, no.~14, pp. 9303--9313, 2021.

\bibitem{ghandehariun2016life}
S.~Ghandehariun and A.~Kumar, ``Life cycle assessment of wind-based hydrogen production in western canada,'' \emph{International Journal of Hydrogen Energy}, vol.~41, no.~22, pp. 9696--9704, 2016.

\bibitem{yue2021hydrogen}
M.~Yue, H.~Lambert, E.~Pahon, R.~Roche, S.~Jemei, and D.~Hissel, ``Hydrogen energy systems: A critical review of technologies, applications, trends and challenges,'' \emph{Renewable and Sustainable Energy Reviews}, vol. 146, p. 111180, 2021.

\bibitem{eiaFAQ}
{U.S. Energy Information Administration}, ``How much of u.s. carbon dioxide emissions are associated with electricity generation?'' \url{https://www.eia.gov/tools/faqs}, 2023, october 2023.

\bibitem{thiruvenkatachari2009post}
R.~Thiruvenkatachari, S.~Su, H.~An, and X.~X. Yu, ``Post combustion co2 capture by carbon fibre monolithic adsorbents,'' \emph{Progress in Energy and Combustion Science}, vol.~35, no.~5, pp. 438--455, 2009.

\bibitem{smith2021cost}
E.~Smith, J.~Morris, H.~Kheshgi, G.~Teletzke, H.~Herzog, and S.~Paltsev, ``The cost of co2 transport and storage in global integrated assessment modeling,'' \emph{International Journal of Greenhouse Gas Control}, vol. 109, p. 103367, 2021.

\bibitem{stehly20212020}
T.~Stehly and P.~Duffy, ``2020 cost of wind energy review,'' National Renewable Energy Lab.(NREL), Golden, CO (United States), Tech. Rep., 2021.

\bibitem{mohammadi2019coordinated}
F.~Mohammadi, M.~Sahraei-Ardakani, Y.~M. Al-Abdullah, and G.~T. Heydt, ``Coordinated scheduling of power generation and water desalination units,'' \emph{IEEE Transactions on Power Systems}, vol.~34, no.~5, pp. 3657--3666, 2019.

\bibitem{goodarzi2022evaluate}
M.~Goodarzi and Q.~Li, ``Evaluate the capacity of electricity-driven water facilities in small communities as virtual energy storage,'' \emph{Applied Energy}, vol. 309, p. 118349, 2022.

\bibitem{goodarzi2021fast}
M.~Goodarzi, D.~Wu, and Q.~Li, ``Fast security evaluation for operation of water distribution systems against extreme conditions,'' in \emph{2021 American Control Conference (ACC)}.\hskip 1em plus 0.5em minus 0.4em\relax IEEE, 2021, pp. 3495--3500.

\bibitem{sedayevatan2023uncertainty}
S.~Sedayevatan, A.~Bahrami, F.~Delfani, and A.~Sohani, ``Uncertainty covered techno-enviro-economic viability evaluation of a solar still water desalination unit using monte carlo approach,'' \emph{Energies}, 2023.

\bibitem{hausmann2021direct}
J.~N. Hausmann, R.~Schl{\"o}gl, P.~W. Menezes, and M.~Driess, ``Is direct seawater splitting economically meaningful?'' \emph{Energy \& Environmental Science}, vol.~14, no.~7, pp. 3679--3685, 2021.

\bibitem{kourkoumpas2016implementation}
D.~Kourkoumpas, E.~Papadimou, K.~Atsonios, S.~Karellas, P.~Grammelis, and E.~Kakaras, ``Implementation of the power to methanol concept by using co2 from lignite power plants: Techno-economic investigation,'' \emph{International journal of hydrogen energy}, vol.~41, no.~38, pp. 16\,674--16\,687, 2016.

\bibitem{Methanolprice}
{CHEMANALYST}, ``{Methanol Price Trend and Forecast},'' \url{https://www.chemanalyst.com/Pricing-data/methanol-1}, accessed on [October 2023].

\bibitem{SEC}
{U.S. Securities and Exchange Commission}, ``{ New Oriental Energy \& Chemical Corporation Provides Update on DME Production Facility Expansion},'' \url{https://www.sec.gov/Archives/edgar/data/1312547}, accessed on [October 2023].

\bibitem{herzog2009carbon}
H.~Herzog, \emph{Carbon dioxide capture and storage}.\hskip 1em plus 0.5em minus 0.4em\relax na, 2009.

\end{thebibliography}

\end{document}